# Summary of the Workshop on:

# Nuclear matrix elements for neutrinoless double beta decay

Institute for Particle Physics Phenomenology
University of Durham, UK

23.-24. 5. 2005

Editor:

Kai Zuber (University of Sussex)

# Introduction

*K.Zuber*

**Abstract**
Neutrinoless double beta decay, which violates total lepton number conservation by two units ($\Delta L = 2$), is of great interest for studying the fundamental properties of neutrinos beyond the standard electroweak theory. High-sensitivity $0\nu\beta\beta$ studies are unique and practical ways for studying the Majorana nature of neutrinos, the $\nu$ mass spectrum, the absolute $\nu$-mass scale, the Majorana CP phases and other fundamental properties of neutrinos in the foreseeable future. On the basis of the recent $\nu$ oscillation studies, the effective mass sensitivity required for observing the $0\nu\beta\beta$ rate is of the order of the atmospheric $\nu$ mass scale of $\delta m_A \sim 50$ meV in the case of an inverted mass hierarchy and of the order of the solar $\nu$ mass scale of $\delta m_S \sim 8$ meV in the case of the normal hierarchy. Theoretical and experimental studies for evaluating nuclear matrix elements are essential for extracting neutrino masses. International coorporative efforts for the determination of $\beta\beta$ matrix elements are encouraged as those for the next-generation $\beta\beta$-experiments. Nuclear matrix elements have to be evaluated with uncertainties of less than 30 % to establish the neutrino mass spectrum. This seems to be possible with the recent improvements in accelerator and detector technology, especially in charge exchange reactions. The neutrino community has defined a number of about a dozen isotopes of potential interest. A coordinated approach to provide experimental information relevant to double beta decay matrix elements for these isotopes is crucial. Experiments include charge exchange nuclear reactions (($^3He, t$), ($d,^2 He$) and others), $\mu$-capture reactions, high-presision mass measurements, $\nu$-nucleus interactions, nucleon transfer reactions and others. The present and planned experiments should be encouraged, and accordingly, continued operation of experimental facilities at RCNP Osaka, KVI Groningen and others are of vital importance since they provide unique opportunities for the experimental studies of the nuclear matrix elements. These facilities should be encouraged to perform the necessary measurements before experience is dooming away and machines are shutdown.

## 1 Motivation

Neutrinos play a fundamental role in various areas of modern physics from particle physics to cosmology. Especially in the case of massive neutrinos, they lead to an extension of the Standard Model of particle physics and might guide the way towards a Grand Unified Theory. Neutrinoless double beta decay plays a crucial role by probing that neutrinos are Majorana particles, determining the absolute neutrinos mass and explorimg possible existing CP-phases associated with the Majorana character, which finally might have an impact on the baryon asymmetry in the Universe via leptogenesis. Recently, all these importance has been summarized in an International Statement for world-wide collaboration on large scale experiments to discover neutrinoless double beta decay. It has also been made clear that nuclear matrix elements are a major uncertainty and their knowledge has to be on a 30 % level to achieve the scientific goals.
The aim of this workshop has been to bring theorists and experimentalists together to review the current status of nuclear matrix element calculations and their input parameters, their shortcomings and precision. Additionally, a goal has been to explore possible experimental measurements to provide those inputs more precisely. The final aim has been to define a coherent strategy of achieving the above mentioned accuracy on nuclear matrix elements to support the experimental searches.

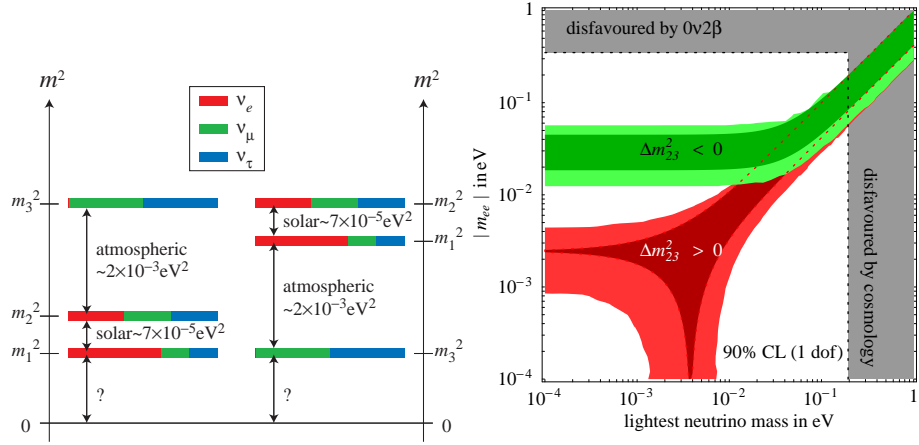

Fig. 1: Left: Possible configurations of neutrino mass states as suggested by oscillations. Currently a normal (left) and an inverted (right) hierarchy cannot be distinguished. The flavour composition is shown as well. Right: The effective Majorana mass $\langle m_{ee} \rangle$ as a function of the lightest mass eigenstate $m_1$. Hierarchical mass patterns can be distinguished for $\langle m_{ee} \rangle$ smaller than 50 meV, otherwise neutrinos can be considered as almost degenerate. Also shown in grey are the regions disfavoured by current $0\nu\beta\beta$-decay limits and a very optimistic limit (could be worse by an order of magnitude) from cosmology (from [2]).

### 1.1 Evidence for a non-vanishing neutrino rest mass - neutrino oscillations

Neutrino physics has been through a revolution over the last ten years, culminating in the recent results of Super-Kamiokande, SNO and KamLAND, see [1]. It is now beyond doubt that neutrinos have a non-vanishing rest mass. However, all the evidence stems from neutrino oscillation experiments, which are not able to measure absolute neutrino masses, but only differences of masses-squared, $\Delta m^2 = m_i^2 - m_j^2$, with $m_i, m_j$ the masses of two neutrino mass eigenstates. To be more specific, in the three neutrino mixing framework the weak eigenstates $\nu_e, \nu_\mu$ and $\nu_\tau$ can be expressed as superpositions of three neutrino mass eigenstates $\nu_1, \nu_2$ and $\nu_3$:

$$\begin{pmatrix} \nu_e \\ \nu_\mu \\ \nu_\tau \end{pmatrix} = \begin{pmatrix} U_{e1} & U_{e2} & U_{e3} \\ U_{\mu 1} & U_{\mu 2} & U_{\mu 3} \\ U_{\tau 1} & U_{\tau 2} & U_{\tau 3} \end{pmatrix} \begin{pmatrix} \nu_1 \\ \nu_2 \\ \nu_3 \end{pmatrix}. \tag{1}$$

To summarize, three pieces of evidence for neutrino oscillations exist:
- The LSND-evidence, $10^{-3} < \sin^2 2\theta < 10^{-1}$, $0.1 eV^2 < \Delta m^2 < 6 eV^2$, $\nu_\mu$- $\nu_e$
- The atmospheric zenith angle dependence observed by Super-Kamiokande and confirmed by K2K, $\sin^2 2\theta$= 1.00, $\Delta m^2 = 2.4 \times 10^{-3} eV^2$, $\nu_\mu$- $\nu_X$
- Solar and reactor neutrinos, $\sin^2 2\theta \approx 0.81$, $\Delta m^2 = 8 \times 10^{-5} eV^2$, $\nu_e$- $\nu_X$

For the sake of simplicity the LSND-evidence will be ignored in the following. To fix the absolute mass scale, direct neutrino mass searches like beta decay and double beta decay are extremely important. Based on the observations, various neutrino mass models have been proposed. These can be categorized as normal hierarchy ($m_3 \gg m_2 \approx m_1$), inverted hierarchy ($m_2 \approx m_1 \gg m_3$) and almost degenerate ($m_3 \approx m_2 \approx m_1$) neutrinos (see Fig. 1).

A benchmark number resulting from the oscillation results is the existence of a neutrino mass eigenstate in the region 10-50 meV, a region only accessible by neutrinoless double beta decay. This scale must be reached to discriminate among the possible mass hierarchies. For the first time in seventy years of double beta decay research there is a prediction with discovery potential. A major input to achieve this goal is a reasonable knowledge of the involved nuclear matrix elements as discussed in the following sections.

## 1.2  Double beta decay

Double beta decay is characterized by a nuclear process changing the nuclear charge Z by two units while leaving the atomic mass A unchanged. It is a transition among isobaric isotopes. All double beta decay emitters are even-even nuclei. Ground states of even-even nuclei have spin 0 and parity (+), thus the transitions are characterized as $(0^+ \to 0^+)$. In addition, also $(0^+ \to 0^+)$ and $(0^+ \to 2^+)$ transitions to excited states can occur.

### 1.21  $2\nu\beta\beta$-decay

Double beta decay was first discussed by M. Goeppert-Mayer in the form of

$$(Z, A) \to (Z + 2, A) + 2e^- + 2\bar{\nu}_e \quad (2\nu\beta\beta\text{-decay}) \tag{2}$$

This process can be seen as two simultaneous neutron decays. This decay mode conserves lepton number and is allowed within the Standard Model, independent of the nature of the neutrino. This mode is of second order Fermi-theory and therefore the lifetime is proportional to $(G_F \cos\theta_C)^{-4}$. As double beta decay is a higher order effect, expected half lives are long compared to $\beta$-decay, rough estimates result in the order of $10^{19-20}$ yrs and higher. The decay rate is given by

$$\lambda_{2\nu}/\ln 2 = (T_{1/2}^{2\nu})^{-1} = G^{2\nu}(Q, Z) \mid M_{GT}^{2\nu} \mid^2 \tag{3}$$

with Q as the transition energy, $G^{2\nu}$ as the phase space and $\mid M_{GT}^{2\nu} \mid^2$ as the involved nuclear matrix elements. They consist of two parts, the transition from the initial nucleus $i$ to an intermediate state $j$ and from there to the daughter nucleus $f$

$$M_{GT}^{2\nu} = \sum_j \frac{\langle 0_f^+ \| t_-\sigma \| 1_j^+ \rangle \langle 1_j^+ \| t_-\sigma \| 0_i^+ \rangle}{E_j + Q/2 + m_e - E_i} \tag{4}$$

with $t_\pm$ and $\sigma$ as isospin- and spin-operators. As two real neutrinos are emitted, the intermediate states $j$ contributing will be $1^+$-states up to several MeV. This decay mode has been observed in about a dozen isotopes.

### 1.22  $0\nu\beta\beta$-decay

Shortly after the classical paper of Majorana discussing a 2-component neutrino, Furry discussed another decay mode in form of

$$(Z, A) \to (Z + 2, A) + 2e^- \quad (0\nu\beta\beta\text{-decay}) \quad . \tag{5}$$

To allow for this process neutrino and antineutrino have to be identical, requiring that neutrinos are Majorana particles. Moreover, to allow for the helicity matching a neutrino mass is required. The total decay rate is then

$$\lambda_{0\nu}/\ln 2 = (T_{1/2}^{0\nu})^{-1} = G^{0\nu}(Q, Z) \mid M_{GT}^{0\nu} + M_T^{0\nu} - \frac{g_V}{g_A} M_F^{0\nu} \mid^2 \left(\frac{\langle m_{ee} \rangle}{m_e}\right)^2 \tag{6}$$

with $M$ as the corresponding Gamow-Teller, tensor and Fermi-contributions. In order to calculate matrix elements in coordinate space, first a neutrino potential has to be evaluated that weakly depends on the energy of the intermediate states and is essentially of a Coulomb-type. The mean value of the momentum of the exchanged neutrino is about $q \propto 1/r$ and thus any intermediate state can be populated. This implies that angular momenta up to $\approx 6\hbar$ have to be considered, ie. forbidden transitions will contribute as well. The quantity of interest, called effective Majorana neutrino mass $\langle m_{ee} \rangle$, is given by

$$\langle m_{ee} \rangle = | \sum_i U_{ei}^2 m_i | = | \sum_i | U_{ei} |^2 \ e^{2i\alpha_i} m_i | \tag{7}$$

with $U_{ei}$ as the mixing matrix elements (see eq. 1), $m_i$ as the corresponding mass eigenvalues and the CP-phases $\alpha_i/2$. The mixing matrix can be separated in two parts, one containing the Majorana phases and one having the same form as the CKM-matrix in the quark sector, here called $U_{MNS}$

$$U = U_{MNS} diag(1, e^{i\alpha_2}, e^{i\alpha_3}) \tag{8}$$

The dependence of $\langle m_{ee} \rangle$ on the lightest mass eigenstate $m_1$ is graphically shown in Fig. 1. Hence, current searches for $0\nu\beta\beta$-decay are driven by two motivations:

1. Verification of the evidence claimed from a $^{76}$Ge measurement [3].
2. If proven not to be true, to build a large scale experiment to reach a sensitivity suggested by the oscillation results, i.e. below 50 meV.

The signature for $0\nu\beta\beta$-decay is a peak at the Q-value of the nuclear transition. The total decay rate scales with $Q^5$ compared to the $Q^{11}$-dependence of $2\nu\beta\beta$-decay.

To summarise, $0\nu\beta\beta$-decay is the gold plated channel to probe the fundamental character of neutrinos, whether being Majorana particles or not. If being Majorana particles, $0\nu\beta\beta$-decay is the only way to probe the two new CP-phases associated with that property, which are not accessible in oscillation experiments. The Majorana character of the neutrino and the CP-phases might play a crucial role in leptogenesis, a process to explain the observed baryon asymmetry of the Universe with the help of lepton number violation. Additional lepton number violating processes can be explored by double beta searches, because every process violating lepton number by two units might contribute. Among them are right-handed (V+A) weak currents, doubled charged higgs bosons and bilinear as well as various trilinear R-parity violating SUSY couplings. All those beautiful conclusions rely on a conversion from a half-life measurement (eq. 6) and thus their final accuracy depends crucially on the knowledge of the involved nuclear matrix elements.

*1.23 Nuclear matrix elements*

The crucial formula for all experimental searches is given by eq. 6. The major source of uncertainty converting an obtained half-life into a neutrino mass is given by the involved nuclear transition matrix elements, for recent reviews see [4, 5, 6, 7]. In contrast to the enormous effort going into the experimental search for $0\nu\beta\beta$-decay there is a lack of coherent effort in determining the nuclear matrix elements more accurately by measuring the relevant quantities. A precision of about 20 % on the nuclear matrix elements is mandatory to guarantee a reasonable conversion from half-lives to neutrino masses keeping the precision high enough to discriminate among the neutrino mass models and determine the neutrino mass. The importance of this was recently highlighted by an International statement of the double beta community. Two types of calculations exist to perform the complex task, nuclear shell model calculations and quasi-particle random phase approximations (QRPA-calculations). Both approaches have strengths and weaknesses, as discussed in a recent review [6]. Shell-model calculations have in the past been restricted to light nuclei around $^{48}$Ca, but the development of large-scale calculations in very large model spaces is beginning to relax these restrictions. For example, the first large scale calculations were performed in spaces with dimensions of the order of $10^8$ for $^{76}$Ge and $^{82}Se$ [8].

These is also a major effort in improving the QRPA calculations. The general situation is shown in Fig. 2. Notice also that different multipoles contribute to $2\nu\beta\beta$-decay and $0\nu\beta\beta$-decay due to the fact that the $2\nu\beta\beta$-decay transition strength is determined by $1^+$ states in the intermediate nuclei, while for $0\nu\beta\beta$-decay other multipoles can be dominant (Fig. 3).

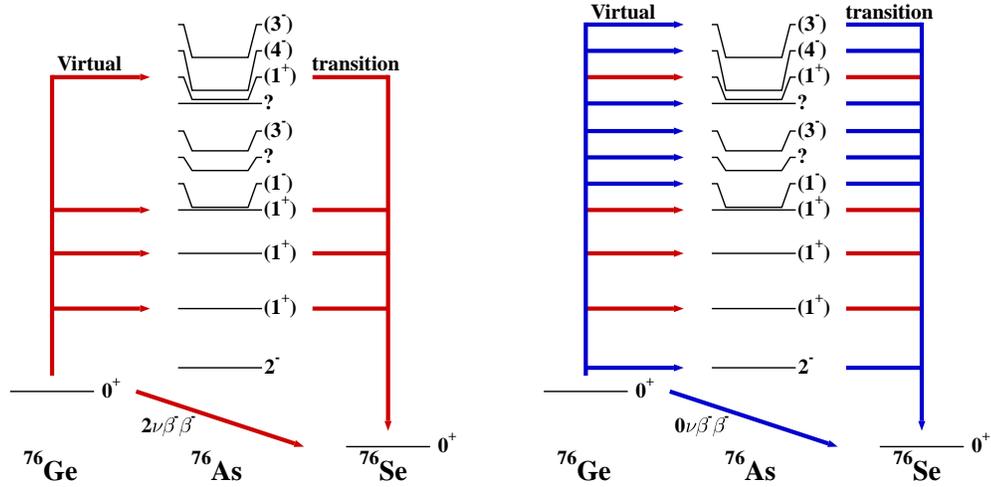

Fig. 2: Schematic view of double beta decay in case of $^{76}$Ge. The double beta transition can be considered as a two step process, from the initial nucleus into an intermediate state ("left leg") followed by the one to the final nucleus ("right leg"). The transition strengths for both transitions have to be measured. Left: The $2\nu\beta\beta$-decay decay scheme which involve only intermediate $1^+$-states. Right: The $0\nu\beta\beta$-decay decay involving all kinds of intermediate states.

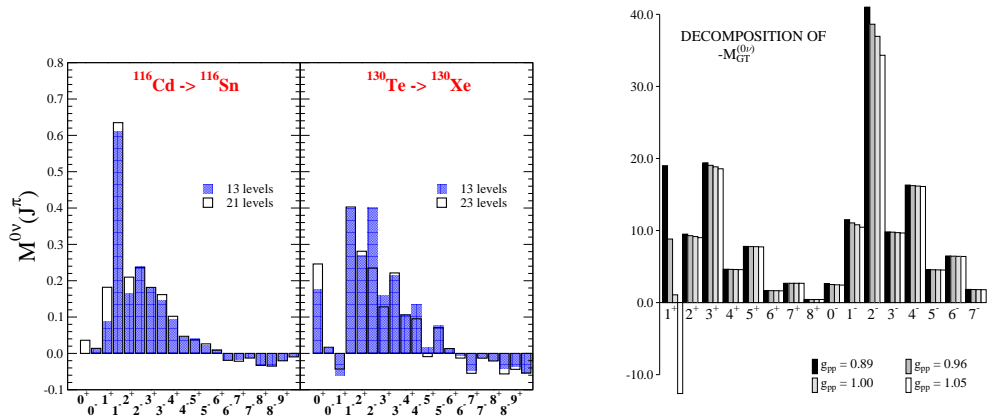

Fig. 3: Calculated contribution of intermediate states to the $0\nu\beta\beta$-decay transition strength according to their momentum and parity assignment $J^\pi$ for various double beta emitters. Left: Calculation for $^{116}$Cd and $^{130}$Te (from [9]). Right: $-M_{GT}^{0\nu}$ for the decay of $^{76}$Ge. Four different values of $g_{pp}$ indicated in the figure, have been included in the analysis (from [10]).

The aim of this workshop has been to define a well planned, coherent strategy to reach this goal, by performing the necessary measurements with currently existing and planned facilities. These measurements should provide reliable input for the theoretical calculations. The outcome of the workshop has been organised in working packages, discussed next.

# Working package 1

*K.Zuber*

## 1 Charge exchange reactions

The obvious way to study and measure GT-transition strengths interesting for double beta decay is charge exchange reactions at accelerators. For an extensive review see [1]. In recent years there has been a significant progress in charge exchange investigations. The advances in detector and accelerator technology presently allow measurements with unprecedented precision. Only because of this, the 30% precision on nuclear matrix elements can be considered feasible at all but requires progress on the same scale as in past years. A coherent measurement of all nine isotopes (see Working package 2) should be performed to measure the low lying $1^+$-transitions relevant for $2\nu\beta\beta$-decay. This will also shed light on the single state dominance (SSD) hypothesis, i.e. the assumption that the lowest lying $1^+$-state is carrying the dominant transition strength in $2\nu\beta\beta$-decay. Tests on SSD can also be performed by measuring single electron spectra in double beta decay experiments using tracking devices like the NEMO-3 experiment. Higher multipole transitions significantly contribute to $0\nu\beta\beta$-decay (see Introduction) and thus serious efforts should be made to establish a reliable procedure to extract information on higher multipoles from double charge-exchange reactions like $(^{11}Be, ^{11}Li)$. The experimental identification of the quantum numbers momentum and parity $J^\pi$ is far from being trivial e.g. it requires polarised beams in charge exchange reactions. There is further the problem, that the correlation between measured cross sections and transition matrix element is complex in hadronic reactions for $\Delta L \neq 0$ and that there exist no relation like the one used for the extraction of the Gamow-Teller strength.

### 1.1 The left leg - Transitions from the mother nucleus to intermediate states

The first part of the double beta transition, namely from the mother isotope to the intermediate state can be studied by single charge exchange reactions like (p,n) or ($^3$He,t). Measurements using the latter process have been performed at RCNP Osaka for $^{100}$Mo [2] and $^{116}$Cd . The state of the art is shown in Fig. 1.

### 1.2 The right leg - Transitions from the intermediate states to daughter nucleus

The second part of information needed can be explored by reactions of the type (n,p), (d,$^2$He) or (t,$^3$He). Such measurements using (d,$^2$He) have recently performed at KVI Groningen for the $^{48}Ti(d,^2He)^{48}Sc$ [3] and the $^{116}$Cd system [4]. The resolution reached here is also shown in Fig. 1.

### 1.3 ft-values for beta decays and electron captures

Additional processes providing important information are beta decay and electron-capture studies. The measurement of their ft-values is of relevance to determine the nuclear matrix elements for the ground state transitions. The importance of such measurements can be discussed for $2\nu\beta\beta$-decay, where the calculations depend on several parameters, the most significant one is the particle-particle coupling strength $g_{pp}$. As it turns out, there is a lively discussion in the community on whether fixing $g_{pp}$ on measured ft-values in beta decay and electron-capture studies or on the value from measured $2\nu\beta\beta$-decayhalf-lives. To clarify the situation it is necessary to perform new ft-value measurements with higher accuracy and for all isotopes of relevance.

**References**

[1] H. Ejiri, *Phys. Rep.* **338**, 265 (2000)

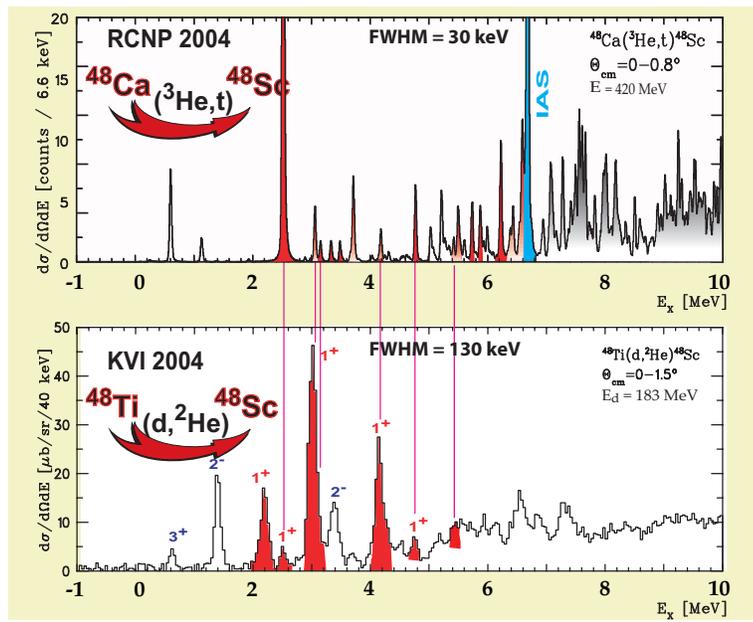

Fig. 1: Measured $^{48}Ca(^{3}He,t)^{48}Sc$ (RCNP Osaka) and $^{48}Ti(d,^{2}He)^{48}Sc$ (KVI Groningen) spectra in charge exchange reactions at 0 degree. Intermediate states which are excited by both reactions are on top of each other.

[2] H. Akimune et al., *Phys. Lett.* B **394**, 23 (1997)

[3] S. Rakers et al., *Phys. Rev.* C **70**, 054302 (2004)

[4] S. Rakers et al., *Phys. Rev.* C **71**, 054313 (2005)

# Working package 2

*K.Zuber*

## 1 Precision mass determinations

There are two reasons for new high-precision mass measurements, namely the precise knowledge of the peak position for $0\nu\beta\beta$-decay and to explore the possible degeneracy of states in the mother and daughter nucleus for double electron-capture, which can lead to a resonant enhancement in the transition rate. The current state-of-the-art technology for high-precision mass determinations is Penning trap mass spectrometry in combination with the time-of-flight cyclotron resonance detection method [1]. Prominent experimental set-ups in this respect are ISOLTRAP at ISOLDE/CERN [3], Geneva and SMILETRAP, Stockholm [4].The latter reaches mass precisions of better than $10^{-9}$. Also the Canadian Penning Trap (CPT) and a new trap under construction in Jyvaskyla (Finland) are in the same sensitivity range. For a recent review see [5].

### 1.1 Q-value determination

The current Q-values for double beta decay deduced from the latest compilation of atomic mass evaluations [2] have a typical uncertainty of about 4 keV. This corresponds to a relative mass precision of about $10^{-7} - 10^{-8}$. The only isotope where the involved masses have already been precisely measured is for $^{76}$Ge decay using SMILETRAP. The resulting Q-value has an uncertainty of only a few ten eV corresponding to a relative precision of $7 \times 10^{-10}$ [6]. This measurement resulted in a factor of 17 improvement in the masses and a factor of 7 in the Q-value. This investigation was motivated by the fact, that for the $^{76}$Ge decay search Ge-semiconductor detectors are used, which have the best energy resolution of all double beta experiments and a precise knowledge of the peak position is crucial.

In the light of next generation experiments, the more demanding challenges and additional other experimental approaches with good energy resolution, it is desirable to extend the set of precision mass determinations to further isotopes. Despite the fact that there are 35 isotopes in nature, for the actual neutrino mass search only those with Q-values beyond 2 MeV are meaningful, because the decay rate scales with $Q^5$. This reduces the number of potentially interesting candidates to 11. Two of them have never been seriously considered due to a lack of a reasonable experimental approach. The remaining nine candidates of interest and their present Q-values including errors, natural abundance and phase space factors are listed in Table 1.1.

Of highest importance are those where the ongoing experiments have good energy resolution. Hence, the two isotopes of main interest would be $^{116}$Cd used by COBRA (CdZnTe semiconductors) and $^{130}$Te used by CUORICINO/CUORE (TeO$_2$ cryogenic bolometers) and COBRA. The precise knowledge of the $^{130}$Te position is also necessary because there is a background line from $^{60}$Co-decay at 2505 keV which might overlap with the $^{130}$Te $0\nu\beta\beta$-decay region. The same argument might hold for $^{136}$Xe searches.

### 1.2 Double electron capture (EC) modes

Recently there has been renewed interest in exploring $EC/EC$ capture modes. As is known for $EC/EC$ capture modes there is a resonance enhancement in the rates if there is an overlap of states in the final and initial nucleus (see Working package 4). A possible system, $^{74}_{34}$Se $\rightarrow^{74}_{34}$Ge, has recently been discussed as an interesting candidate [8]. However, a systematic search might reveal more candidates. One of those is $^{106}$Cd were several states are close to the Q-value of the transition. A decision on the degeneracy can only be made if the Q-values are known to the level of 200 eV. Thus, for finding nuclear levels allowing for resonant enhancement, the masses of the involved isotopes have to been known with a precision of better than $10^{-9}$. This is well within reach of the SMILETRAP Penning trap mass spectrometer and can also be addressed at ISOLTRAP.

Table 1: Compilation of $\beta^-\beta^-$-emitters with a Q-value of at least 2 MeV. Given are Q-values with errors deduced from [2], natural abundances and phase-space factors for $0\nu\beta\beta$-decay and $2\nu\beta\beta$-decay (taken from [7]).

| Transition | Q-value (keV) | nat. ab. (%) | $[G^{0\nu}]^{-1}$ (yr·$eV^2$) | $[G^{2\nu}]^{-1}$ (yr) |
|---|---|---|---|---|
| $^{48}_{20}$Ca→$^{48}_{22}$Ti | 4274 ± 4 | 0.187 | 4.10E24 | 2.52E16 |
| $^{76}_{32}$Ge→$^{76}_{34}$Se | 2039.00 ± 0.05 | 7.8 | 4.09E25 | 7.66E18 |
| $^{82}_{34}$Se→$^{82}_{36}$Kr | 2995.5 ± 1.9 | 9.2 | 9.27E24 | 9.27E24 |
| $^{96}_{40}$Zr→$^{96}_{42}$Mo | 3347.7 ± 2.2 | 2.8 | 4.46E24 | 5.19E16 |
| $^{100}_{42}$Mo→$^{100}_{44}$Ru | 3035 ± 6 | 9.6 | 5.70E24 | 1.06E17 |
| $^{110}_{46}$Pd→$^{110}_{48}$Cd | 2004 ± 11 | 11.8 | 1.86E25 | 2.51E18 |
| $^{116}_{48}$Cd→$^{116}_{50}$Sn | 2809 ± 4 | 7.5 | 5.28E24 | 5.28E24 |
| $^{124}_{50}$Sn→$^{124}_{52}$Te | 2287.8 ± 1.5 | 5.64 | 9.48E24 | 5.93E17 |
| $^{130}_{52}$Te→$^{130}_{54}$Xe | 2530.3 ± 2.0 | 34.5 | 5.89E24 | 2.08E17 |
| $^{136}_{54}$Xe→$^{136}_{56}$Ba | 2462 ± 7 | 8.9 | 5.52E24 | 2.07E17 |
| $^{150}_{60}$Nd→$^{150}_{62}$Sm | 3367.7 ± 2.2 | 5.6 | 1.25E24 | 8.41E15 |

# Working package 3

*K.Zuber*

## 1 Muon capture on nuclei

A process suggested in [1] to explore nuclear matrix elements for double beta decay is ordinary muon capture (OMC)

$$\mu^- + (A, Z) \rightarrow (A, Z-1) + \nu_\mu \tag{1}$$

As is obvious in OMC the right leg of double beta decay will be explored. Due to the large mass of the muon, captures to higher multipolarities than $1^+$ are not forbidden in the sense of electron capture and $2^-, 3^+$, etc. states are easily populated by OMC. The muon capture rate $W$ is given by

$$W = 4P(\alpha Z m'_\mu)^3 \frac{2J_f + 1}{2J_i + 1}(1 - \frac{q}{m_\mu + AM})q^2 \tag{2}$$

with $A$ being the mass of the initial and final nuclei, $Z$ the charge of the initial nucleus and $m'_\mu$ the reduced muon mass. Furthermore, $\alpha$ denotes the fine-structure constant, $M$ the average nucleon mass and $q$ the magnitude of the exchanged momentum between the captured muon and the nucleus. The term $P$ contains the nuclear matrix elements involved and it is here, where in the limit of $q, Z \rightarrow 0$ the element $M[101]$ can be linked to $M_{GT}^{2\nu}$ if the final state is a $1^+$ as shown in [2].

Experimentally muon capture on nuclei is a well established technique, however first attempts on double beta decay elements have only recently been started at PSI [3]. The systems explored are the Ar and Ca decays. Especially in case of $^{48}$Ca, it is the capture process on the daughter, $^{48}$Ti, which has to be explored. The nuclei measured are still relatively light to be described by shell model calculations. However, the comparison between theoretical contribution and experimental results indicates that still some better understanding is necessary (fig. 1). It has been shown that the total capture rates are dominated by the GT resonance where negative parity dipole states are dominating [4] and give a general handle on the matrix elements. More information could be obtained by exploring partial capture rates which for heavier double beta isotopes becomes more complex because of more complex nuclear level schemes. In the determination of the partial muon capture rates the effect of feeding of the lower lying states by gamma cascades and of possible open channels for particle emission from the higher states in the final nucleus have to be corrected for. This could be done with gamma detectors or arrays having large detection efficiencies. Several such arrays exist in the nuclear spectroscopy community (EXOGAM, GAMMASPHERE, MINIBALL, EUROBALL) and there are plans for arrays of new generation (AGATA). Additionally, the assignment of quantum numbers to the various nuclear levels is important and has to be supported by other measurements like those described in working package 1. Facilities are available at PSI, RIKEN-RAL, TRIUMF and the future J-PARC facility. From statistical arguments DC-machines are preferred with respect to pulsed machines. Beam requirements would be about a week per target allowing for another week to set-up the experiment. The list of the nine most important isotopes mentioned in Working package 2 and a few possible $\beta^+\beta^+$-candidates could thus be measured in about 3 month beam-time.

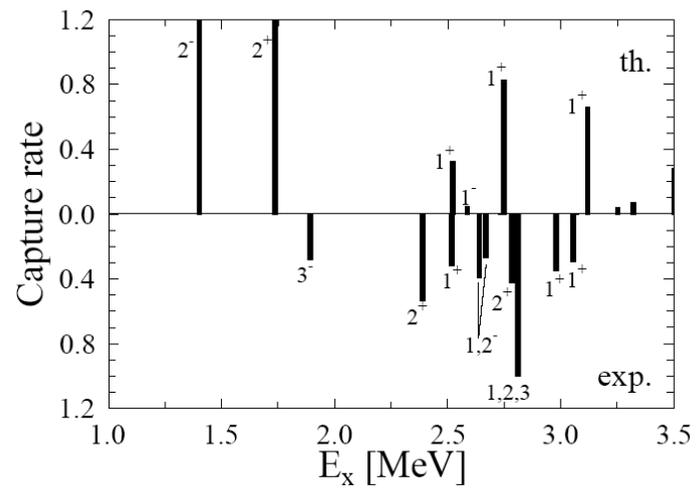

Fig. 1: Multipole composition of muon capture on $^{48}$Ti as measured and calculated.

# Working package 4

*K.Zuber*

## 1 Double electron-capture and $\beta^+\beta^+$-modes

Three different decay channels can be considered here

$$\begin{align}
(Z,A) &\to (Z-2,A) + 2e^+ + (2\nu_e) & (\beta^+\beta^+ - decay) \\
e^- + (Z,A) &\to (Z-2,A) + e^+ + (2\nu_e) & (\beta^+/EC - decay) \\
2e^- + (Z,A) &\to (Z-2,A) + (2\nu_e) & (EC/EC - decay)
\end{align} \quad (1)$$

where the last two cases involve electron-capture (EC). Especially the $\beta^+/EC$ mode shows an enhanced sensitivity to right handed weak currents [1] und thus provides a valueable information on the physics involved if is observed. The experimental signatures of the decay modes involving positrons in the final state are promising because of two or four 511 keV photons. Despite this nice signature, they are less often discussed in literature, because for each generated positron the available Q-value is reduced by 2 $m_e c^2$, which leads to much smaller decay rates than in comparable $\beta^-\beta^-$ -decays. Hence, for $\beta^+\beta^+$ -decay to occur, the Q-values must be at least 2048 keV. Only six isotopes are know to have such a high Q-value. The full Q-value is only available in the $EC/EC$ mode. Its detection is experimentally more challenging, basically requiring the concept of source equal to detector to observe the produced X-rays or Auger electrons.

Recently the resonant enhancement of decay rates in the radiative $0\nu EC/EC$ has been considered in case that nuclear levels in the mother and daugther nuclei are at least almost degenerate [2]. In case the Q-value of the transition is close to the 2P-1S atomic level difference (within about 1 keV), this allows for a resonant enhancement. However, to explore this resonance effect more precisely, Q-value measurements and thus precision mass determinations (see Working Package 2) must be done. An accuracy much better than 1 keV is required.

Currently $EC/EC$ isotopes used in running experiments are $^{106}$Cd (experiments COBRA and TGV2) and $^{64}$Zn , $^{108}$Cd , $^{120}$Te (COBRA).

**References**

[1] M. Hirsch et al., *Z. Phys.* C **347**, 151 (1994)

[2] Z. Sujkowski, S. Wycech, *Phys. Rev.* C **70**, 052501 (2004)

# Working package 5

*K.Zuber*

## 1 (Anti-)Neutrino - Nucleus scattering

The idea of using neutrino-nucleus interactions as a constrain for the nuclear matrix elements involved in neutrinoless double-beta decay has been proposed recently, and discussed as a future option in the context of next generation neutrino beams from accelerators [1]. Besides high intensity conventional neutrino beams (superbeams) and neutrino factories also the concept of beta beams has been brought forward for neutrino oscillation studies. It is meant to be an ion accelerator for beta unstable isotopes. Their decay would result in flavour pure $\bar{\nu}_e$ or $\nu_e$ beams, depending on whether $\beta^-$- or $\beta^+$-isotopes are stored. A low energy version of such a beta-beam facility [2] and a different size (smaller) for the considered storage ring [3] would be suitable for nuclear matrix elements explorations [1]. Building such a storage ring with a nearby detector would allow to study neutrino-nucleus interaction and hence transition matrix elements in greater detail, because varying the average neutrino beam energy $\langle E_\nu \rangle$ would result in populating different intermediate states and thus different multipoles. This is shown in fig. 1. $\langle E_\nu \rangle$ is determined by $E_\nu \sim 2\gamma Q_\beta/2$ where there relativistic Lorentz factor $\gamma$ can be varied. Both legs of double beta transitions can be explored by using $\nu_e$ and $\bar{\nu}_e$ beams. On the other hand, because of their high intensities, conventional beams offer a very interesting option as well, allowing to explore one of the two legs only.

**References**

[1] C. Volpe, *Journal of Physics G* **31**, 903 (2005)

[2] C. Volpe, *Journal of Physics G* **30**, L1 (2004)

[3] J. Serreau, C. Volpe, *Phys. Rev.* C **70**, 055502 (2004)

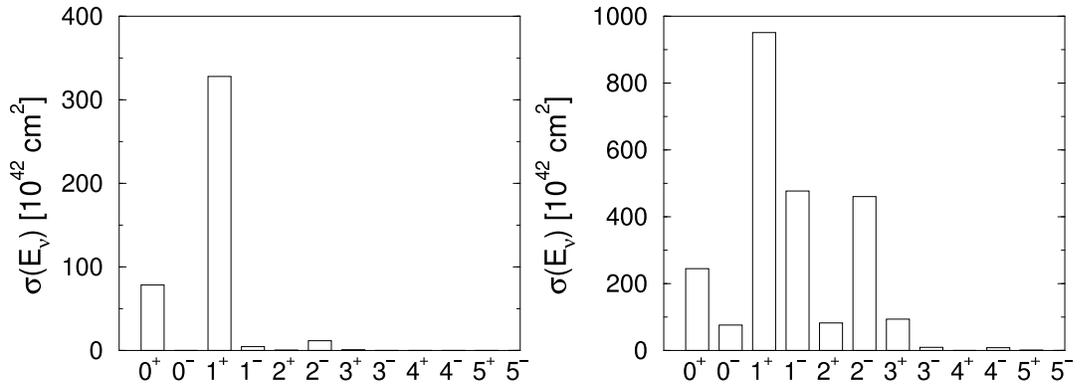

Fig. 1: Multipole decomposition for neutrino-nucleus scattering with $\langle E_\nu \rangle$ = 30 MeV (left) and $\langle E_\nu \rangle$= 60 MeV (right). Clearly visible is the different contribution of states to both scattering cross sections (from [1]).

# Working package 6

*K.Zuber*

## 1 Nucleon transfer reactions

Shell-model calculations of nuclear matrix elements involve detailed calculations of the wavefunctions of the participating nuclei. There are several ways in which relevant aspects of the calculated wavefunctions may be tested against experimental measurements of other observables. For example, the determination of the relative occupation of the valence orbitals that contribute to the ground-state wavefunctions can be performed using single-nucleon transfer reactions on the relevant targets. Two-nucleon transfer reactions on even-even systems, where transfers connecting the $0^+$ ground states are enhanced and reflect pairing correlations, should probe aspects of the wavefunctions which are important for double beta decay where a zero-coupled pair of neutrons is converted to a zero-coupled pair of protons.

Comparisons of measured cross sections with those corresponding to the calculated wavefunctions will help constrain the calculations. Some information on one-nucleon transfer is already available from earlier studies, but no comprehensive study was undertaken. Such a general programme benefits from a consistent approach where common procedures are adopted in acquisition, analysis and interpretation of data. Facilities for such studies are currently available and work is already underway by an Argonne-Berkeley-GANIL-Manchester-Yale collaboration to make precise and consistent measurements of nuclei relevant to $^{76}$Ge double beta decay.

# Summary

*K.Zuber*

Neutrinoless double beta decay is a crucial process in neutrino physics for many reasons as discussed in the introduction. There is an enormous experimental effort to observe this decay and prove that neutrinos are Majorana particles. However, the measured quantity is a half-life and to extract a reasonable neutrino mass the involved nuclear matrix elements have to be known at least on a level of only 30 % uncertainty.

The major aim of this workshop has been to identify the major uncertainties in the ingredient for nuclear matrix elements calculations, explore possible strategies for improvement and define a coherent effort between theory and experiment to reduce the current uncertainties finally to a precision to 30 %, which is needed for a reliable neutrino mass determination. It has been decided to focus on nine out of thirty five existing $\beta^-\beta^-$-isotopes, those which are the most promising and sensitive ones for neutrinoless double beta decay searches.

Three groups of experimental processes were identified, charge exchange reactions, ordinary muon capture and neutrino-nucleus scattering. The priority of the measurements fits nicely with the timeline of the available facilities to perform these. Of highest priority are charge exchange measurements (working package 1), which can be done at facilities like KVI Groningen and RCNP Osaka. It is strongly recommended to use the facility at KVI Groningen to perform the important measurements as discussed, because both measurements are complementary and necessary. Additional measurements from nucleon transfer reactions (working package 6) will help to determine the wavefunctions of the involved nuclei for shell model calculations. The next additional experimental piece of information becoming available is ordinary muon capture (working package 3). First attempts have started at PSI using the lightest double beta system $^{48}$Ca - $^{48}$Ti , but results are still inconclusive. The study of muon capture on heavy nuclei and the extraction of the relevant information has to be explored in greater detail. Last but not least, neutrino nucleus scattering at various energies could be used (working package 5). For that low energy beta beams, neutrino beams produced by radioactive decays in flight, are an ideal tool. Intensive investigations on the feasibility of such a facility are presently ongoing. Very intensive conventional sources could also offer an interesting option.

Slightly decoupled from the nuclear matrix elements but important as well is a precise knowledge of the actual transitions energy, i.e. the Q-value to search for the neutrinoless double beta peak (working package 2). Penning trap mass spectrometers are predestined for that, they can improve the precision by more than an order of magnitude for almost all isotopes. Additionally, this is of major importance in the field of the complementary process of double electron capture, interest in that has revived recently (working package 4). If there is a degeneracy between states in the mother and daughter atom, resonance enhancement of the decay is expected, however requiring a precise knowledge of the energy levels. First interesting systems have been found.

Complementary to the above mentioned time-line several other important points have been discussed during the meeting which deserve further investigations:

- Nuclear matrix element calculations depend on various parameters, among the most significant is the particle-particle coupling $g_{pp}$. To compare $g_{pp}$ values deduced from $2\nu\beta\beta$-decay and those from electron-capture and beta-decay, new measurements of ft-values in electron-capture and beta-decay should be performed to resolve this important point (working package 3).

- Many nuclei undergoing double beta decay are deformed and it is of large interest to study the effect of deformation on the $0\nu\beta\beta$-decay matrix element calculations. Until now, QRPA calculations for $0\nu\beta\beta$-decay matrix elements were performed by assuming the spherical symmetry. Recently, the $2\nu\beta\beta$-decay transitions were studied within deformed QRPA. It has been found that matrix elements are suppressed with respect to spherical case. More precisely, a sizable reduction effect that scales with the deformation difference between parent and daughter nuclei has been recognised. Thus, there is a strong interest to investigate this effect also in the calculation of the $0\nu\beta\beta$-decay matrix elements. For that purpose precise measurements of nuclear deformations are stronlgy required. QRPA calculations normally do not include deformation of nuclei. Those should be measured, if not done, and be implemented in the code.

- The validity of the single state dominance assumption has to be explored in larger detail.
- $N\pi$ and $\Delta$-exchange contribution mix in at the 10-20 % level, making a measurement of double charge exchange reactions desirable.
- Electromagnetic transitions are an additional tool for study because they proceed via the isobaric analogue state. This can be linked to double beta decay transitions.

# Acknowledgements

It is a pleasure to thank the IPPP Durham for its hospitality, especially Prof. James Stirling and Mrs. Linda Wilkinson.

# Participants

- A. Barabash (Moscow)
- K. Blaum (Mainz)
- H. Ejiri (Osaka)
- S. Freeman (Manchester)
- M. Freer (Birmingham)
- D. Frekers (Münster)
- K. Jungmann (KVI Groningen)
- R. Saakyan (UC London)
- F. Simkovic (Bratislava)
- J. Suhonen (Jyvaskyla)
- J. Stirling (Durham)
- C. Volpe (Orsay)
- H. Woertche (KVI Groningen)
- S. Wychech (Warsaw)
- K. Zuber (Sussex)